\documentclass[aps,pre,twocolumn,amssymb,superscriptaddress]{revtex4}
\usepackage{subfigure}
\usepackage{graphicx}
\usepackage{rotating} 
\usepackage{color}
\usepackage{lscape}
\usepackage{rotating}
\usepackage{amsmath,amsthm}
\usepackage{mathtools}
 
\usepackage{epstopdf}
\begin{document}
\title{Physical insight into the thermodynamic uncertainty relation using Brownian motion in tilted periodic potentials}

\author{Changbong Hyeon}
\thanks{hyeoncb@kias.re.kr}
\affiliation{Korea Institute for Advanced Study, Seoul 02455, Korea}

\author{Wonseok Hwang}
%\thanks{}
\affiliation{Korea Institute for Advanced Study, Seoul 02455, Korea}

\begin{abstract}
Using Brownian motion in periodic potentials $V(x)$ tilted by a force $f$, we provide physical insight into the thermodynamic uncertainty relation, a recently conjectured principle for statistical errors and irreversible heat dissipation in nonequilibrium steady states. According to the relation, nonequilibrium output generated from  dissipative processes necessarily incurs an energetic cost or heat dissipation $q$, and in order to limit the output fluctuation within a relative uncertainty $\epsilon$, at least $2k_BT/\epsilon^2$ of heat must be dissipated. 
Our model shows that this bound is attained not only at near-equilibrium ($f\ll V'(x)$) but also at far-from-equilibrium $(f\gg V'(x))$, more generally when the dissipated heat is normally distributed.  
Furthermore, the energetic cost is maximized near the critical force when the barrier separating the potential wells is about to vanish and the fluctuation of Brownian particle is maximized. 
These findings indicate that the deviation of heat distribution from Gaussianity gives rise to the inequality of the uncertainty relation, further clarifying the meaning of the uncertainty relation.   
Our derivation of the uncertainty relation also recognizes a new bound of nonequilibrium fluctuations that the variance of dissipated heat ($\sigma_q^2$) increases with its mean ($\mu_q$) and cannot be smaller than $2k_BT\mu_q$. 
\end{abstract}

\maketitle 

Precise determination of an output information from a thermodynamically dissipative process necessarily incurs energetic cost to generate it.       
Trade-offs between energetic cost and information processing in biochemical and biomolecular processes have been highlighted for the last decades  \cite{barato2015PRL,barato2015JPCB,Lan2012NaturePhysics,Parrondo2015NaturePhysics,Hinczewski2014PRX}. 
Among others, Barato and Seifert \cite{barato2015PRL} have recently conjectured a fundamental bound in the minimal heat dissipation ($q$) to generate an output with relative uncertainty %for a given fluctuation inherent to output observable 
($\epsilon$). 
% for cyclic processes epitomized with enzyme reactions and molecular motors. 
To be specific, when a molecular motor moves along cytoskeletal filament \cite{Visscher99Nature,Kolomeisky07ARPC,Hyeon11BJ}, the chemical free energy transduced into the motor movement, which results in a travel distance of the motor $X(t)$, is eventually dissipated as heat into the surrounding media \cite{Bustamante2005PhysicsToday,Hwang2017JPCL}, the amount of which increases with the time ($\langle q\rangle\sim t$). 
                                                                                                                                                                                                                                                                                                           %, $q\sim T\sigma t$, where $T$ is the ambient temperature and $\sigma$ is the entropy production rate. 
Because of the inherent stochasticity of chemical processes, 
the travel distance $X(t)$ has its own variance $\sigma_X^2=\langle(\delta X(t))^2\rangle$, and defines a time-dependent fluctuation in the output $\epsilon_X\left[\equiv  \sigma_X/\mu_X\right]$, whose squared quantity decreases with time $t$, as $\epsilon_X^2\sim t^{-1}$.   
The product of the two quantities, $\mathcal{Q}$, %=T\sigma t\epsilon_X^2$ 
is, in fact, independent of $t$ \cite{barato2015PRL,pietzonka2017finite}, and it was further argued that $\mathcal{Q}$ is always greater than $2k_BT$ for any process that can be described as Markov jump process on a suitable network. 
This notion is concisely written as 
\begin{align}
\mathcal{Q}=\langle q\rangle\times\epsilon_X^2\geq 2k_BT. 
\label{eqn:inequality}
\end{align}    
The validity of this inequality was claimed for general Markovian networks \cite{barato2015PRL,barato2015JPCB}, and was partly proved at near equilibrium, linear response regime \cite{barato2015PRL}.  
This effort has recently been followed by a general proof employing the large deviation theory \cite{Gingrich2016RPL,pietzonka2016PRE,Polettini2016PRE}. 

Here, while limited to a particular model, we provide a less abstract and physically more tangible proof of the thermodynamic uncertainty relation (Eq.\ref{eqn:inequality}) than the existing studies by considering dynamics of a Brownian particle on a static periodic potential subjected to a nonconservative force $f$. 
Projection of biomolecular processes in 1D periodic potentials is fully legitimate as long as the time scale separation between a slow variable of interest and other faster variables is ensured \cite{ZwanzigBook}, and the Brownian motion in 1D periodic potential has routinely been employed in describing the motion of molecular motors and enzyme turnover reactions \cite{Kolomeisky07ARPC,astumian2010BJ,Hwang2017JPCL}. 
We directly calculate each term ($\langle q\rangle$ and $\epsilon_X^2$) in Eq.\ref{eqn:inequality}, and show that the product of the two quantities must be greater than $2k_BT$.

The overdamped Langevin equation for the ``position" $x(t)$ of a quasi-particle on a periodic potential $V(x)=V(x+L)$ is written as 
\begin{align}
%\dot{x}(t)&=\mu(-V'(x(t))+f)+\eta(t)\nonumber\\
\dot{x}(t)=\mu F(x(t),f)+\eta(t)
\label{eqn:Langevin}
\end{align} 
where $\mu$ is the motility coefficient (or inverse of friction coefficient $\gamma$, $\mu=\gamma^{-1}$), $F(x(t),f)\equiv -V'(x(t))+f$, and Gaussian white noise is assumed such that $\langle \eta(t)\rangle=0$ and $\langle \eta(t)\eta(t')\rangle=2D\delta(t-t')$ with the diffusion constant $D=\mu k_BT$. 
Then, the corresponding Fokker-Planck equation for the probability density $\rho(x,t)$ is 
\begin{align}
\partial_t\rho(x,t)&=D\partial_x[\partial_x-\beta F(x(t))]\rho(x,t)\nonumber\\
&=-\partial_xj(x,t)
\label{eqn:FK}
\end{align}
which defines the probability current $j(x,t)=-D\partial_x\rho(x,t)+\mu F(x(t))\rho(x,t)$. 

Eqs. \ref{eqn:Langevin} and \ref{eqn:FK} represent the Brownian motion in tilted washboard potentials where the extent of tilt is controlled by $f$ \cite{Risken}. 
At steady state, the probability density becomes time independent ($\rho(x,t)=\rho_{\text{ss}}(x)$), which in turn renders a constant probability current, $j(x,t)=j_{\text{ss}}$. 
Furthermore, heat (housekeeping heat \cite{sekimoto1998langevin,Hatano01PRL,Seifert2012RPP}) has to be continuously dissipated to sustain the process at nonequilibrium steady state (NESS).  
Before carrying out an explicit calculation, two limiting cases are worth considering: 
(i) At weak tilt ($0\lesssim f\ll V'(x)$), it is expected that the particle hops stochastically between the adjacent confining potential wells of $V(x)$ with a mean hopping time $\langle\tau\rangle$ and a small downhill velocity $\sim L/\langle\tau\rangle$ and a reduced effective diffusion constant $D_{\text{eff}}(<D)$ \cite{LifsonJCP1960,ZwanzigPNAS88}.   
(ii) At strong tilt ($f\gg V'(x)$), the barrier separating the adjacent potentials vanishes, the particles will drift downhill with mean steady state velocity $v_{\text{ss}}\sim\mu f$ and with the original diffusion constant $D$. 

Here, our aim is to evaluate $\mathcal{Q}$ for arbitrary value of $f$ and prove that $\mathcal{Q}$ is lower bounded by $2k_BT$.  
First, the heat dissipated from this process in NESS is a housekeeping heat, which can be evaluated for  Langevin systems as \cite{Hatano01PRL,Seifert2012RPP,barato2015PRL}. 
\begin{align}
q[x(t),f]=\mu^{-1}\int_0^td\tau v_{\text{ss}}(x,f)\dot{x}(\tau)
\end{align}
where 
$v_{\text{ss}}(x,f)\equiv j_{\text{ss}}(f)/\rho_{\text{ss}}(x,f)=\mu F(x,f)-D\partial_x\log{\rho_{\text{ss}}(x,f)}$
is the mean local velocity.  
 Note that due to the stochastic term $\dot{x}(\tau)$, the housekeeping heat is a stochastic quantity that depends on the path of realization.  
The argument $[x(t)]$ of $q$ makes it explicit that the calculated heat is for a particular realization of the trajectory $[x(t)]=(x(0),x(1),x(2),\ldots,x(t))$, hence the housekeeping heat averaged over the ensemble of trajectories is written in the following form.  
\begin{align}
\langle q\rangle&=\int dx_0p(x_0)\int \prod_{i=1}^{t-1}dx_iP(x_i|x_{i-1})q(x_0,\cdots x_t)\nonumber\\
&=\int\mathcal{D}[x(\tau)]e^{-\mathcal{S}[x(\tau)|x_0]}\int_0^td\tau \mu^{-1}v_{\text{ss}}(x(\tau))\dot{x}(\tau)\nonumber\\
&=\Big\langle \int_0^td\tau v_{\text{ss}}(x(\tau)) F(x(\tau))\Big\rangle\nonumber\\
&=\langle v_{\text{ss}}(x,f)F(x,f)\rangle t\equiv \langle vF\rangle_{\text{ss}}t
\label{eqn:heat}
\end{align}
where $\mathcal{S}[x(\tau),f(\tau)]=\int_0^td\tau\left(\frac{(\dot{x}-\mu F(x,f))^2}{4D}+\frac{\mu}{2}\partial_xF\right)$ \cite{Seifert2012RPP}, and $\langle\ldots\rangle\equiv \int\mathcal{D}[x(\tau)]e^{-\mathcal{S}[x(\tau)|x_0]}(\ldots)$ denotes the average over all the paths and initial conditions. 
At steady state, however, the implicit time dependence in $v_{\text{ss}}(x(\tau))$ and $F(x(\tau))$ can be removed from the above formal path integral expression, and thus the notion of sum over all paths is replaced with an integral weighed with the steady state probability, $\rho_{\text{ss}}(x)$ with a normalization condition, $\int_0^L\rho_{\text{ss}}(x)dx=1$. 
That is, 
$\langle g\rangle =\int \mathcal{D}(x(\tau))P[x(\tau)|x_0]\int_0^td\tau g(x(\tau))=\int dxg(x)\rho_{\text{ss}}(x)$. 
In the last line of Eq.\ref{eqn:heat}, we have dropped the dependence of steady-state quantities on $x$ and $f$ from the expression and used a simplified notation $\langle\cdots\rangle_{\text{ss}}$. 
Henceforth, for notational convenience, we will use this simplified notation, i.e., $\langle K\rangle_{\text{ss}}\equiv \langle K_{\text{ss}}(x,f)\rangle=\int_0^LK_{\text{ss}}(x,f)\rho_{\text{ss}}(x,f)dx$.

Next, the mean travel distance $\langle X(t)\rangle$ and its variance ($\langle(\delta X(t))^2\rangle$) of Brownian motion in tilted periodic potentials are the topic that has been discussed in many different contexts, and their analytic forms at $t\rightarrow \infty$ are available from Ref. \cite{Reimann2001PRL,Reimann2002PRE,wang2011effective}. 
\begin{align}
\langle X(t)\rangle&=\mu\Big\langle\int_0^td\tau F[x(\tau),f]\Big\rangle=\mu\langle F\rangle_{\text{ss}} t
%&=t\times\mu \int_0^L(\gamma j_{\text{ss}}+k_BT\partial_x\rho_{\text{ss}}(x))dx\nonumber\\
\label{eqn:mean}
\end{align}
and
\begin{align}
\langle(\delta X(t))^2\rangle=2D_{\text{eff}}(f) t
\label{eqn:variance}
\end{align}
where $D_{\text{eff}}(f)=D\mathcal{G}(f)$ is a force dependent effective diffusion coefficient. 
\begin{align}
\mathcal{G}(f)=\frac{\langle I_{\mp}(x,f)I_+(x,f)I_-(x,f)\rangle_L}{\langle I_{\mp}(x,f)\rangle^3_L}, 
\end{align}
here, $\langle \cdots\rangle_L\equiv L^{-1}\int_0^L(\cdots)dx$ denotes averaging over a period with
\begin{align}
I_+(x,f)=e^{\beta\Phi(x,f)}\int_{x-L}^{x}dye^{-\beta \Phi(y,f)}
\end{align}
and
\begin{align}
I_-(x,f)=e^{-\beta\Phi(x,f)}\int_{x}^{x+L}dye^{\beta \Phi(y,f)}. 
\end{align}
where $\Phi(x,f)\equiv V(x)-fx$.

Now, we are ready to evaluate $\mathcal{Q}$ (Eq.\ref{eqn:inequality}) using Eqs. \ref{eqn:heat}, \ref{eqn:mean}, \ref{eqn:variance}, and prove the uncertainty relation, $\mathcal{Q}\geq 2k_BT$.  
\begin{align}
\mathcal{Q}&=\langle q\rangle\times \frac{\langle(\delta X(t))^2\rangle}{\langle X(t)\rangle^2}\nonumber\\
%&=t\langle v_{\text{ss}}(x,f)F(x,f)\rangle\frac{2Dt}{t^2\mu^2\langle F(x,f)\rangle^2}\nonumber\\
&=2k_BT\frac{\langle vF\rangle_{\text{ss}}}{\mu\langle F\rangle_{\text{ss}}^2}\mathcal{G}(f), 
\label{eqn:Q}
\end{align}
The two core averages in Eq.\ref{eqn:Q} are evaluated as follows. 
\begin{align}
\langle F\rangle_{\text{ss}}&=\int_0^LF(x)\rho_{\text{ss}}(x)dx\nonumber\\
&=\int_0^L(\gamma j_{\text{ss}}(f)+k_BT\partial_x\rho_{\text{ss}}(x,f))dx=\gamma j_{\text{ss}}(f)L
\label{eqn:F}
\end{align} 
and 
\begin{align}
\langle vF\rangle_{\text{ss}}&=\int_0^Lv_{\text{ss}}(x)F(x)\rho_{\text{ss}}(x)dx\nonumber\\
&=\int_0^Lv_{\text{ss}}(x,f)(\gamma j_{\text{ss}}(f)+k_BT\partial_x\rho_{\text{ss}}(x,f))dx\nonumber\\
%&=\int_0^L(\gamma j_{\text{ss}}^2(f)\rho_{\text{ss}}^{-1}(x,f)+\beta^{-1}j_{\text{ss}}(f)\partial_x\log{\rho_{\text{ss}}(x,f)})dx\nonumber\\
&=\gamma j^2_{\text{ss}}(f)\int_0^L\rho_{\text{ss}}^{-1}(x,f)dx
\label{eqn:vF}
\end{align}
where the condition of periodic boundary $\rho_{\text{ss}}(L)=\rho_{\text{ss}}(0)$ was used in both Eqs.~\ref{eqn:F} and \ref{eqn:vF}. 
Now, to prove the uncertainty relation, we have to verify the following inequality for all $f$. 
\begin{align}
\frac{\mathcal{Q}(f)}{2k_BT}
&=\left(\frac{1}{L^2}\int_0^L\rho_{\text{ss}}^{-1}(x,f)dx\right)\times \mathcal{G}(f)\geq 1.  
\end{align}

The conditions of normalization $\int_0^L\rho_{\text{ss}}(x,f)dx=1$ and boundedness $|\rho_{\text{ss}}(x,f)|<\infty$ for all $x$'s
enable us to calculate the position dependent steady state probability $\rho_{\text{ss}}(x,f)$  
 and steady state current $j_{\text{ss}}(f)$ \cite{Risken} (see Appendix A), both of which are required for evaluating $\mathcal{Q}$ explicitly.  
\begin{align}
&\rho_{\text{ss}}(x,f)\nonumber\\
&=\frac{j_{\text{ss}}(f)e^{-\beta\Phi(x,f)}}{D\Omega(\beta fL)}\left(\psi_+(L,f)-\Omega(\beta fL)\psi_+(x,f)\right), 
\label{eqn:rho}
\end{align}
\begin{align}
j_{\text{ss}}(f)=\frac{D\Omega(\beta fL)}{\psi_+(L,f)\psi_-(L,f)- \Omega(\beta fL)\Psi_+(L,f)},
\label{eqn:j}
\end{align}
where $\Psi_+(L,f)=\int_0^Ldxe^{-\beta\Phi(x,f)}\psi_+(x,f)$, $\beta=1/k_BT$, $\Omega(x)=1-e^{-x}$, and $\psi_{\pm}(x,f)=\int_0^xe^{\pm\beta\Phi(x',f)}dx'$. 
Insertion of Eqs.\ref{eqn:rho} and \ref{eqn:j} into $\langle F\rangle_{\text{ss}}$ (Eq.~\ref{eqn:F}) and $\langle vF\rangle_{\text{ss}}$ (Eq.~\ref{eqn:vF}) allows us to calculate $f$-dependence of $\mathcal{Q}$, which is depicted in Fig.1. 

\begin{figure}
	\centering
	\includegraphics[width=1\linewidth]{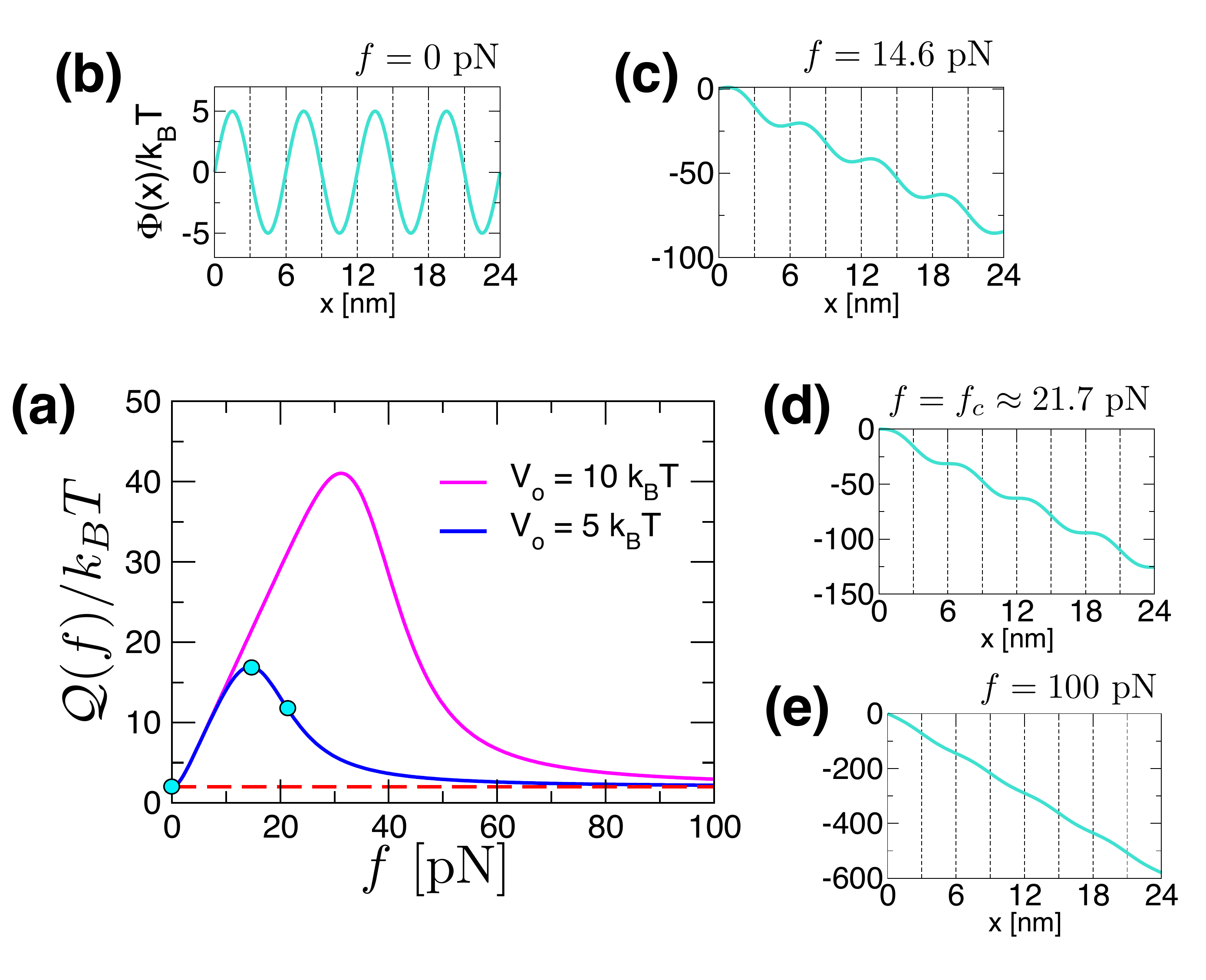}
	\caption{$\mathcal{Q}(f)$ calculated for $\Phi(x,f)=V(x)-fx$ with $V(x)=V_o\sin{(2\pi x/L)}$, $V_o=5$ $k_BT$, $L=6$ nm (blue line in panel (a)). 
	The shape of potential $\Phi(x,f)$ at varying $f$ are shown in the small panels (b)-(e). 
	A greater dissipation, (larger $\mathcal{Q}(f)$) is obtained for a deeper potential well ($V_o=10$ $k_BT$, magenta line). 
	The red dashed line depicts the lower bound $\mathcal{Q}=2k_BT$. 
	}
	\label{CA_1_vs_10}
\end{figure}

Three regimes of $f$ are investigated: (i) In the weak tilt limit ($V'(x)\gg f\rightarrow 0$), using Eqs.~\ref{eqn:F}, \ref{eqn:vF}, \ref{eqn:rho}, \ref{eqn:j}, and the periodicity of $V(x)$, i.e, $\int_{x-L}^xe^{\pm\beta V(y)}dy=\int_0^Le^{\pm\beta V(y)}dy$, one can show that  
$\langle F\rangle_{\text{ss}}\rightarrow f\varphi_+^{-1}\varphi_-^{-1}$ where $\varphi_{\pm}\equiv\langle e^{\pm\beta V(x)}\rangle_L$
, $\langle vF\rangle_{\text{ss}}\rightarrow \mu f^2\varphi_+^{-1}\varphi_-^{-1}$, and 
$\mathcal{G}(f)\rightarrow \varphi_+^{-1}\varphi_-^{-1}$, 
where $\varphi_{\pm}=\langle e^{\pm\beta V(x)}\rangle_L$
This leads to $\mathcal{Q}(f)/2k_BT\rightarrow 1$. 
A series expansion of $\mathcal{Q}(f)$ in the power of $f$ leads to (see Appendix B-1 for the details of power series expansion)
\begin{align}
\lim_{f\rightarrow 0\ll V'(x)}\frac{\mathcal{Q}}{2k_BT}=1+\lambda \beta f+\mathcal{O}(f^2)\geq 1 
\label{eqn:limit1}
\end{align}
with $\lambda\geq 0$ (Eq.S25); 
(ii) In the strong tilt limit ($V'(x)\ll f$), the particle continuously drifts downhill without being trapped in potential wells. In this case, 
$\langle F\rangle_{\text{ss}}\rightarrow f$, 
$\langle vF\rangle_{\text{ss}}\rightarrow \mu f^2$, and 
$\mathcal{G}(f)\rightarrow 1$, which again leads to $\mathcal{Q}(f)/2k_BT\rightarrow 1$. 
A series expansion of $\mathcal{Q}(f)$ in the power of $1/f$ leads to (See Appendix B-2)
\begin{align}
\lim_{V'(x)\ll f}\frac{\mathcal{Q}}{2k_BT}\geq 1+3\frac{\langle \{V'(x)\}^2\rangle_L}{f^3}+\cdots\geq 1.  
\label{eqn:limit2} 
\end{align}
(iii) Fig.1 shows that $Q(f)$ is maximized at an intermediate tilt limit $f\lesssim f_c$ which satisfies $V'(x)-f_c\approx 0$. In this case, $\Phi(x,f)$ resembles a descending staircase (see Fig.1 at $f\approx f_c$). 
It is straightforward to show that (see Appendix B-3) 
\begin{align}
\frac{Q(f_c)}{2k_BT}&\simeq\frac{\beta f_cL}{3}\frac{\left(1+5e^{-\beta f_cL}+5e^{-2\beta f_cL}+e^{-3\beta f_cL}\right)}{(1+e^{-\beta f_cL})^2(1-e^{-\beta f_cL})}\nonumber\\
&\geq 1, 
\end{align}
where $\mathcal{Q}(f_c)$ increases monotonically with $f_c$ and the equality is attained only for $f_cL\ll k_BT$, which is realized essentially for a barrierless, flat potential. 

Taking (i), (ii), (iii) together, we have essentially proved that $\mathcal{Q}(f)/2k_BT\geq 1$ for all $f$. Remarkably, the lower bound of $\mathcal{Q}$, $2k_BT$, is attained at the two disparate  conditions of $f$, and it is of particular note that $\mathcal{Q}$ is maximized near the critical force ($f\lesssim f_c$) at which the barrier of potential is about to vanish and the fluctuation of Brownian particle is maximized.

Instead of the travel distance $X(t)$ as an output observable, 
the dissipated heat ($q$) in steady state can be used as an alternative probe of output from the process.  
For a given thermodynamic affinity per cycle $\mathcal{A}$, which itself is a deterministic quantity defined as the log-ratio between the forward and backward flux (or microscopic rate constants of chemical networks) \cite{barato2015PRL,Hwang2017JPCL,Qian07ARPC}, the mean and variance of the housekeeping heat can be related with those of another stochastic observable, such as reaction cycle step $n(t)$ or travel distance $X(t)$:  
$\langle q\rangle=\mathcal{A}\times \langle n\rangle=\mathcal{A}/L\times \langle X\rangle$ and $\langle(\delta q)^2\rangle=\mathcal{A}^2\times \langle (\delta n)^2\rangle=(\mathcal{A}/L)^2\times \langle (\delta X)^2\rangle$. 
Then, the squared uncertainty $\epsilon^2_{\xi}=\sigma^2_{\xi}/\mu_{\xi}^2$ are identical for $\xi=q(t)$, $n(t)$, $X(t)$. 

If the dissipated heat itself is used as an output observable ($\xi=q$), 
$\mathcal{Q}$ turns into Fano factor of $q$ and is lower bounded by $2k_BT$.  
\begin{align}
\mathcal{Q}=\mu_q\times \epsilon_{q}^2=\sigma^2_q/\mu_q\geq 2k_BT,   
\label{eqn:Q_q}
\end{align}
equivalently,  
\begin{align}
0\leq \mu_q \leq \frac{\sigma_q^2}{2k_BT}. 
\label{eqn:new_inequality}
\end{align}
The lower bound of $\mu_q$ is a straightforward outcome of the 2nd law of the thermodynamics, but the upper bound, which is an interesting outcome, implies that the variance of heat dissipated in NESS is constrained by its mean, such that $\sigma_q^2$ grows with $\mu_q$ and cannot be smaller than $2k_BT\times \mu_q$. 
While the thermodynamic uncertainty relation has originally recognized the bound of 
steady-state current fluctuation ($\epsilon_J$) \cite{Gingrich2016RPL}, 
here we show that the relation of variance of heat dissipation with its mean (Eq.\ref{eqn:new_inequality}) can be deduced from it as well.  

Lastly, we relate Eqs.\ref{eqn:Q_q} and \ref{eqn:new_inequality} with the integral fluctuation theorem for housekeeping heat, $\langle e^{-\beta q}\rangle=1$ \cite{speck2005JPA}. 
The first inequality in Eq.\ref{eqn:new_inequality} ($0\leq \mu_q$) is obtained from Jensen's inequality for convex functions, $\langle e^{-\beta q}\rangle \geq e^{-\beta \langle q\rangle}$, 
and the second one ($\mu_q\leq \sigma_1^2/2k_BT$) is obtained by truncating the cumulant expansion of $e^{-\beta q}$ at the second term as
\begin{align}
0&=-\beta^{-1}\log{\langle e^{-\beta q}\rangle}\nonumber\\
&=\mu_q -\frac{\beta}{2}\sigma_q^2+\mathcal{R}(\beta q) \geq \mu_q -\frac{\beta}{2}\sigma^2_q, 
\label{eqn:IFT}
\end{align}
and by claiming that the remaining sum of alternating series beyond the second cumulant is non-negative regardless of the value of $\beta q$, i.e., $\mathcal{R}(\beta q)\geq 0$. In order to be consistent with the form of thermodynamic uncertainty relation in Eq.\ref{eqn:Q_q}, the inequality $\mathcal{R}(\beta q)\geq 0$ should hold.   
The equality sign is acquired when the heat distribution, $P(q)$, is Gaussian. 

Speck \emph{et al.} \cite{speck2005JPA}, in fact, have calculated $P(q,t)$ by solving the Fokker-Planck equation of Brownian particle in periodic potentials through adiabatic elimination of fast variables. They found that the dissipated heat in steady states takes a form of Gaussian distribution $P(q,t)\sim \exp{\left[-(q-\mu_q)^2/2\sigma^2_q\right]}$ at two limiting cases.  
For $|V'(x)|\gg f$
\begin{align}
P(q,t)\sim \exp{\left[-\frac{(q-\gamma \langle v^2\rangle_{\text{ss}} t)^2}{4\gamma k_BT\langle v^2\rangle_{\text{ss}} t}\right]}. 
\end{align}
which gives $\mu_q=\gamma\langle v^2\rangle_{\text{ss}}t$, $\sigma_q^2=2\gamma k_BT\langle v^2\rangle_{\text{ss}}t$, and $\mathcal{Q}=\sigma^2_q/\mu_q=2k_BT$. 
For $|V'(x)|\ll f$,   
\begin{align}
P(q,t)\sim \exp{\left[-\frac{(q-\mu f^2t)^2}{4Df^2t}\right]}
\end{align}
which gives $\mu_q=\mu f^2 t$, $\sigma^2_q=2Df^2t$. In both cases, $\mathcal{Q}=\sigma^2_q/\mu_q=2k_BT$.  
These results (i) agree with those shown in Eqs.\ref{eqn:limit1} and \ref{eqn:limit2} that use the travel distance as an output ($\mathcal{Q}=\mu_q\epsilon_X^2=2k_BT$), (ii) confirm that $\mathcal{Q}$ is time-independent, and (iii) that Gaussian dissipation leads to $\mathcal{Q}=2k_BT$ as we have discussed using the integral fluctuation theorem (Eq.\ref{eqn:IFT}). 

Lastly, the implication of the thermodynamic uncertainty relation, which differs from other inequality relations such as the second law and stability condition in equilibrium thermodynamics \cite{CallenBook}, is worth further deliberation.  
It is interesting that the lower bound of $\mathcal{Q}$ is attained both at near and far from equilibrium. 
For unicyclic Markovian network with $N$ states, it was shown that $\mathcal{Q}\geq (\mathcal{A}/N)\coth{(\mathcal{A}/2Nk_BT)}\geq 2k_BT$ \cite{barato2015PRL}. 
In this case, the minimum dissipation bound of $\mathcal{Q}=2k_BT$ is attained only for the affinity goes to zero ($\mathcal{A}\rightarrow 0$), in which the local detailed balance condition is approached. 
Our limiting case of $f\gg V'(x)$ could be thought of as $\mathcal{A}\gg 1$, but the above expression of $\mathcal{Q}$ from Markovian network diverges as $\mathcal{A}\gg 1$ at fixed $N$, which appears to contradict with Eq.\ref{eqn:limit2}. 
Under the condition of $f\gg V'(x)$, however, the potential barriers between adjacent wells vanish and the mapping of dissipative dynamics using Markov jump process on networks can no longer be maintained. 
While the condition is fundamentally different from another limiting case near equilibrium, we still find the $\mathcal{Q}=2k_BT$ bound being attained at this extreme driving condition that also gives rise to a Gaussian heat dissipation, a hallmark of independence and uncorrelated statistics. 
Also suggested in Fig.1, except for the two limiting conditions, it is likely that the distribution of dissipated heat is characterized with non-Gaussianity.   
Thus, $\mathcal{Q}$, greater than $2k_BT$, implies deviation of heat dissipation from Gaussian and can be used as a measure for assessing the complexity of dissipative processes which, for the case of Brownian motion in 1D periodic potentials, approaches to its lower bound both at near and far from equilibrium. 
\\

\section*{ACKNOWLEDGEMENTS}
We thank Fyl Pincus and Hyunggyu Park for valuable insight and illuminating discussions. 
\\

\appendix
\setcounter{equation}{0}
\renewcommand{\theequation}{A\arabic{equation}}

\section{Derivation of $\rho_{ss}(x,f)$ and $j_{ss}(f)$ \cite{Risken}.} 
The Fokker-Planck equation with effective potential $\Phi(x,f)=V(x)-fx$ is  
\begin{align}
\partial_t \rho(x,f,t)&=D\partial_x(\partial_x+\beta\Phi'(x;f))\rho(x,t)\nonumber\\
&=-\partial_xj(x,f,t). 
\end{align}
At steady state, $\rho(x,f,t)=\rho_{ss}(x,f)$ and $j(x,f,t)=j_{\text{ss}}(f)$,  
\begin{align}
j_{ss}(f)&=-D\partial_x\rho_{ss}(x,f)-\mu\Phi'(x)\rho_{ss}(x,f)\nonumber\\
&=-De^{-\beta\Phi(x,f)}\partial_x(e^{\beta\Phi(x,f)}\rho_{ss}(x,f))
\end{align}
Here, $\rho_{\text{ss}}(x,f)$ can be formally solved as 
\begin{align}
\rho_{ss}(x,f)&=Ne^{-\beta\Phi(x,f)}\nonumber\\
&-\frac{j_{ss}(f)}{D}e^{-\beta\Phi(x,f)}\int^xe^{\beta\Phi(y,f)}dy
\end{align}
and one can determine $N$, the normalization constant, by using the fact that the steady state probability $\rho_{ss}(x)$ is bounded for large $x$, and $j_{ss}$ by using $\int_0^L\rho_{ss}(x)dx=1$. 

To carry out this algebra, let us first evaluate 
\begin{widetext}
\begin{align}
\int_0^{x+nL}e^{\beta\Phi(x')}dx'&=\int_0^Le^{\beta\Phi(x')}dx'+\cdots+\int_{(n-1)L}^{nL}e^{\beta\Phi(x')}dx'+\int_{nL}^{nL+x}e^{\beta\Phi(x')dx'}\nonumber\\
&=\left(1+e^{-\beta fL}+\cdots+e^{-\beta f(n-1)}\right)\int_0^Le^{\beta\Phi(x')}dx'+e^{-\beta fnL}\int_0^xe^{\beta\Phi(x')}dx'\nonumber\\
&=\frac{1-e^{-\beta fnL}}{1-e^{-\beta fL}}I+e^{-\beta fnL}\int_0^xdx'e^{\beta V(x')}
\end{align}
where $I\equiv\int_0^Le^{\beta\Phi(x')}dx'$. 
This allows us to express $\rho_{ss}(x+nL)$ as  
\begin{align}
\rho_{ss}(x&+nL)=e^{-\beta\Phi(x+nL)}\left(N-\frac{j_{ss}}{D}\int_0^{x+nL}e^{\beta\Phi(x')}dx'\right)\nonumber\\
&=e^{-\beta\Phi(x,f)}e^{\beta fnL}\left[N-\frac{j_{ss}}{D}\left(I\frac{1-e^{-\beta fnL}}{1-e^{-\beta fL}}+e^{-\beta fnL}\int_0^xdx'e^{\beta V(x')}\right)\right]\nonumber\\
&=e^{-\beta\Phi(x,f)}e^{\beta fnL}\left[N-\frac{j_{ss}I}{D(1-e^{-\beta fL})}\right]+e^{-\beta \Phi(x,f)}\left(\frac{j_{ss}I}{D}\frac{1}{1-e^{-\beta fL}}-\frac{j_{ss}}{D}\int_0^xdx'e^{\beta V(x')}\right)
\label{eqn:rho_ss}
\end{align}
\end{widetext}
In order for $\rho_{ss}(x+nL)$ to be bounded even when $n\rightarrow \infty$ ($f>0$), the first term in the last line of Eq.\ref{eqn:rho_ss} should vanish, which demands
\begin{align}
N=\frac{j_{ss}I}{D(1-e^{-\beta fL})}
\end{align}
Therefore, the steady state probability along the reaction coordinate is written as  
\begin{align}
\rho_{ss}(x)=\frac{j_{ss}}{D}e^{-\beta\Phi(x,f)}\left(\frac{\psi_+(L)}{1-e^{-\beta fL}}-\psi_+(x)\right)
\label{eqn:rho}
\end{align}
where $\psi_{\pm}(x)=\int_0^xe^{\pm\beta\Phi(x')}dx'$. 

Next, the normalization condition $\int_0^L\rho_{ss}(x,f)dx=1$ determines $j_{ss}(f)$ 
\begin{align}
&j_{ss}(f)\nonumber\\
&=\frac{D(1-e^{-\beta fL})}{\psi_+(L,f)\psi_-(L,f)-(1-e^{-\beta fL})\int_0^Ldxe^{-\beta\Phi(x,f)}\psi_+(x,f)}. 
\end{align}
%%%%%%%%%%%%%%%%%%%%%%%%%%%%%%%%%%%%%%%%%%%%%%%

\renewcommand{\theequation}{B\arabic{equation}}
\section{Behaviors of $\mathcal{Q}(f)$ at three regimes of $f$ \cite{Reimann2001PRL,Reimann2002PRE,wang2011effective}}
For three different regimes of $f$, we will evaluate 
\begin{align}
\frac{\mathcal{Q}(f)}{2k_BT}=\frac{1}{L^2}\int_0^Ldx\rho_{\text{ss}}(x,f)\times \frac{\langle (I_-(x,f))^2I_+(x,f)\rangle_L}{\langle I_-(x,f)\rangle_L^3}.
\label{eqn:Q_ratio}
\end{align}
To evaluate Eq.\ref{eqn:Q_ratio}, it is convenient to have the following 4 identities \cite{Reimann2001PRL,Reimann2002PRE,wang2011effective}. 
First, 
\begin{align}
I_+(x,f)&=e^{\beta\Phi(x,f)}\int_{x-L}^xdye^{-\beta\Phi(y,f)}\nonumber\\
&=\int_0^Ldye^{\beta(\Phi(x,f)-\Phi(x-y,f))}\nonumber\\
&=\int_0^Ldye^{\beta(V(x)-V(x-y))}e^{-\beta fy}
\end{align}
Second, 
\begin{align}
I_-(x,f)&=e^{-\beta\Phi(x,f)}\int_x^{x+L}dye^{\beta\Phi(y,f)}\nonumber\\
&=\int_0^Ldye^{-\beta(\Phi(x,f)-\Phi(x+y,f))}\nonumber\\
&=\int_0^Ldye^{-\beta(V(x)-V(x+y))}e^{-\beta fy}
\end{align}
%%%%%%%%%%%%%%%%%
%\item $\frac{1}{L}\int_0^Ldx\int_0^Ldyye^{-\beta (V(x)+V(x-y))}$
Third, 
\begin{align}
&\frac{1}{L}\int_0^Ldx\int_0^Ldyye^{-\beta (V(x)+V(x-y))}\nonumber\\
&=\frac{1}{L}\int_0^Ldx\int_0^Ldy(L-y)e^{-\beta (V(x)+V(x+y))}\nonumber\\
&=L\varphi_-^2-\frac{1}{L}\int_0^Ldx\int_0^Ldyye^{-\beta (V(x)+V(x+y))}\nonumber\\
&=L\varphi_-^2-\frac{1}{L}\int_0^Ldx\int_0^Ldyye^{-\beta (V(x-y)+V(x))}, 
\end{align}
which leads to 
\begin{align}
\frac{1}{L}\int_0^Ldx\int_0^Ldyye^{-\beta (V(x-y)+V(x))}=\frac{L\varphi_-^2}{2}. 
\label{eqn:integral}
\end{align}
%\vspace{1cm}
%\item $l_o^2\equiv \frac{1}{L}\int_0^Ldxe^{-\beta V(x)}\int_0^Ldyye^{\beta V(x+y)}$.
Lastly, 
\begin{align}
l_o^2&=\frac{1}{L}\int_0^Ldxe^{-\beta V(x)}\underbrace{\int_0^Ldyye^{\beta V(x+y)}}_{\int_x^{L+x}ds(s-x)e^{\beta V(s)}}\nonumber\\
&=\frac{1}{L}\int_0^Ldxe^{-\beta V(x)}\left(L\langle x\rangle_++L\psi_+(x,0)-xL\varphi_+\right)\nonumber\\
&=L\langle x\rangle_+\varphi_--L\langle x\rangle_-\varphi_++\Psi_+(L,0)
\label{eqn:l12}
\end{align}
where $\langle x^n\rangle_{\pm}\equiv \frac{1}{L}\int_0^Ldxx^ne^{\pm\beta V(x)}$ and 
$\Psi_+(x,f)=\int_0^xdye^{-(\beta V(y)-fy)}\int_0^ydze^{\beta (V(z)-fz)}$, thus 
$\Psi_+(L,0)=\int_0^Ldxe^{-\beta V(x)}\int_0^xdxe^{\beta V(x)}$. 
%\end{itemize}
%\vspace{1cm}
%\item $\Psi_+(L,0)$
%\begin{align}
%&\Psi_+(L,0)=\int_0^Ldxe^{\beta V(x)}\int_0^xdye^{-\beta V(y)}\nonumber\\
%&=\int_0^Ldxe^{\beta V(x)}\left(\int^L_0dye^{-\beta V(y)}-\int_x^Ldye^{-\beta V(y)}\right)\nonumber\\
%&=L^2\varphi_+\varphi_--\int_0^Ldxe^{\beta V(x)}\int_x^Ldye^{-\beta V(y)}
%\label{eqn:Psi}
%\end{align}
\vspace{1cm}

\subsection{Weak tilt limit, $V'(x)\gg f$}
First, 
\begin{align}
I_-(x,f)&=\int_0^Ldye^{-\beta(V(x)-V(x+y))}(1-\beta fy+\cdots)\nonumber\\
&=e^{-\beta V(x)}\int_0^Ldye^{\beta V(y)}\nonumber\\
&-\beta f\int_0^Ldyye^{-\beta(V(x)-V(x+y))}+\mathcal{O}(f^2)
\end{align} 
where we have used a property of periodic function, $\int_0^Ldye^{\beta V(x+y)}=\int_0^Ldye^{\beta V(y)}$. 
Then, the average over a period is  
\begin{align}
\langle &I_-(x,f)\rangle_L=\frac{1}{L}\underbrace{\int_0^Ldx e^{-\beta V(x)}}_{=L\varphi_-}\underbrace{\int_{0}^Ldye^{\beta V(y)}}_{=L\varphi_+}\nonumber\\
&-\beta f\underbrace{\frac{1}{L}\int_0^Ldx\int_0^Ldyye^{-\beta(V(x)-V(x+y))}}_{=l_o^2}+\mathcal{O}(f^2)\nonumber\\
&=L\varphi_+\varphi_-\left(1-\frac{l_o^2\beta f}{L\varphi_+\varphi_-}+\mathcal{O}(f^2)\right)
\end{align}
where $\varphi_{\pm}=\frac{1}{L}\int_0^Ldxe^{\pm\beta V(x)}$ and $l_o^2=\frac{1}{L}\int_0^Ldxe^{-\beta V(x)}\int_0^Ldyye^{\beta V(x+y)}$. 
Hence, we obtain 
\begin{align}
\langle I_-(x,f)\rangle_L^3&=L^3\varphi_+^3\varphi_-^3\left(1-\frac{3l_o^2}{L\varphi_+\varphi_-}\beta f+\mathcal{O}(f^2)\right)
\end{align}

Next, 
\begin{widetext}
\begin{align}
\langle (I_-(x,f))^2I_+(x,f)\rangle_L&=\frac{1}{L}\int_0^Ldx(I_-(x))^2\left[\int_0^Ldye^{\beta(V(x)-V(x-y))}e^{-\beta fy}\right]\nonumber\\
&=\frac{1}{L}\int_0^Ldxe^{-\beta V(x)}\left(L^2\varphi_+^2-2L\varphi_+\int_0^Ldyye^{\beta V(x+y)}\beta f+\cdots\right)\nonumber\\
&\qquad\qquad\qquad\qquad \times\left(\underbrace{\int_0^Ldye^{-\beta V(x-y)}}_{=L\varphi_-}-\beta f\int_0^Ldyye^{-\beta V(x-y)}+\cdots\right)\nonumber\\
&=\underbrace{\frac{1}{L}\int_0^Ldxe^{-\beta V(x)}}_{\varphi_-}L^3\varphi_+^2\varphi_-
-2\beta fL^2\varphi_+\varphi_-\underbrace{\frac{1}{L}\int_0^Ldxe^{-\beta V(x)}\int_0^Ldyye^{\beta V(x+y)}}_{\equiv l_o^2}\nonumber\\
&\qquad\qquad\qquad\qquad\qquad\qquad-\beta fL^2\varphi_+^2\underbrace{\frac{1}{L}\int_0^Ldxe^{-\beta V(x)}\int_0^Ldyye^{-\beta V(x-y)}}_{=L^2\varphi_-^2/2 \text{ (Eq.\ref{eqn:integral})}}
\nonumber\\
%&=L^3\varphi_+^2\varphi_-^2-\beta f(2L^2\varphi_+\varphi_-l_1^2+L^4\varphi_+^2\varphi_-^2/2)+\mathcal{O}(f^2)\nonumber\\
&=L^3\varphi_+^2\varphi_-^2\left(1-\frac{2l_o^2+L^2\varphi_+\varphi_-/2}{L\varphi_+\varphi_-}\beta f+\mathcal{O}(f^2)\right)
\end{align}
Therefore, 
\begin{align}
&\frac{\langle (I_-(x,f))^2I_+(x,f)\rangle_L}{\langle I_-(x,f)\rangle_L^3}=\frac{1}{\varphi_+\varphi_-}\left[1+\left(\frac{l_o^2}{L^2\varphi_+\varphi_-}-\frac{1}{2}\right)\beta fL+\mathcal{O}(f^2)\right]
\end{align}

Lastly, 
\begin{align}
\frac{1}{L^2}\int_0^L\rho_{\text{ss}}^{-1}(x,f)dx
&=\frac{1}{L^2}\int_0^Ldx\frac{e^{\beta\Phi(x,f)}[\psi_+(L,f)\psi_-(L,f)-\Omega(\beta fL)\Psi_+(x,f)]}{\psi_+(L,f)-\Omega(\beta fL)\psi_+(x,f)}\nonumber\\
&=\frac{1}{L^2}\int_0^Ldxe^{\beta\Phi(x,f)}\frac{\mathcal{N}(x,f)}{\mathcal{D}(x,f)}
\label{B:rho}
\end{align}
where $\Psi_+(x,f)=\int_0^xdye^{-\beta \Phi(y,f)}\psi_+(y,f)=\int_0^xdye^{-\beta \Phi(y,f)}\int_0^ydze^{\beta\Phi(z,f)}$. For small $f$ ($f\ll V'(x)$), 
\begin{align}
\psi_{\pm}(L,f)&=\int_0^Ldxe^{\pm\beta\Phi(x,f)}=\int_0^Ldxe^{\pm\beta V(x)}(1\mp\beta fx+\cdots)\nonumber\\
&=L\varphi_{\pm}\mp(\beta fL)\langle x\rangle_{\pm}+\cdots
\end{align}
with $\langle x^n\rangle_{\pm}=\frac{1}{L}\int_0^Ldxx^ne^{\pm\beta V(x)}$. 
Thus, one can expand the numerator and denominator defined respectively as $\mathcal{N}(x,f)$ and $\mathcal{D}(x,f)$ in Eq.\ref{B:rho} in the power of $f$:
\begin{align}
\mathcal{N}(x,f)&=\psi_+(L,f)\psi_-(L,f)-\Omega(\beta fL)\Psi_+(x,f)\nonumber\\
&=\left(L\varphi_+-(\beta fL)\langle x\rangle_{+}+\cdots\right)\left(L\varphi_-+(\beta fL)\langle x\rangle_{-}+\cdots \right)-\Omega(\beta fL)\Psi_+(x,f)\nonumber\\
&=L^2\varphi_+\varphi_--\left(L\langle x\rangle_{+}\varphi_--L\langle x\rangle_{-}\varphi_++\Psi_+(x,0)\right)(\beta f L)+\mathcal{O}(f^2)
\label{B:N}
\end{align}
and 
\begin{align}
\mathcal{D}(x,f)&=\psi_+(L,f)-\Omega(\beta fL)\psi_+(x,f)\nonumber\\
%&=\left(L\varphi_++L\sum_{n=1}\frac{1}{n!}(\beta f)^n\langle x^n\rangle_{+}\right)-\Omega(\beta fL)\psi_+(x,f)\nonumber\\
&=L\varphi_+-(\langle x\rangle_++\psi_+(x,0))\beta fL+\mathcal{O}(f^2). 
\label{B:D}
\end{align}

Eqs. \ref{B:rho}, \ref{B:N}, and \ref{B:D} lead to the power series expansion of $\frac{1}{L^2}\int_0^L\rho^{-1}_{\text{ss}}(x,f)$ in terms of $f$   
\begin{align}
\frac{1}{L^2}\int_0^Ldx&\rho_{\text{ss}}^{-1}(x,f)
=\frac{1}{L^2}\int_0^Ldxe^{\beta\Phi(x,f)}\frac{\mathcal{N}(x,f)}{\mathcal{D}(x,f)}\nonumber\\
&=\frac{1}{L^2}\int_0^Ldxe^{\beta V(x)}\left(1-\beta fx+\cdots \right)\frac{L^2\varphi_+\varphi_--\left(L\langle x\rangle_{+}\varphi_--L\langle x\rangle_{-}\varphi_++\Psi_+(x,0)\right)(\beta f L)+\cdots}{L\varphi_+-[\langle x\rangle_++\psi_+(x,0)]\beta fL+\cdots}\nonumber\\
%&\approx \frac{1}{L}\int_0^Ldxe^{\beta V(x)}\left(1-\beta fx+\cdots \right)\nonumber\\
%&\qquad\qquad\times\left(\varphi_--\frac{\left(\langle x\rangle_{+}\varphi_--\langle x\rangle_{-}\varphi_++\Psi_+(x,0)/L\right)}{\varphi_+}\beta f+\cdots\right)\left(1+\frac{\langle x\rangle_++\psi_+(x,0)}{\varphi_+}\beta f+\cdots \right)\nonumber\\
&\approx \varphi_+\varphi_-+\left[\langle x\rangle_-\varphi_+-\langle x\rangle_+\varphi_-+\frac{1}{L\varphi_+}\int_0^Ldxe^{\beta V(x)}\left\{\varphi_-\psi_+(x,0)-\frac{\Psi_+(x,0)}{L}\right\}\right]\beta f+\mathcal{O}(f^2)\nonumber\\
&=\varphi_+\varphi_-\left[1+\left(\frac{\langle x\rangle_-}{\varphi_-}-\frac{\langle x\rangle_+}{\varphi_+}+\frac{1}{\varphi_+}\int_0^Ldxe^{\beta V(x)}\left\{\frac{\psi_+(x,0)}{L\varphi_+}-\frac{\Psi_+(x,0)}{L^2\varphi_+\varphi_-}\right\}\right)\beta f+\mathcal{O}(f^2)\right]. 
\end{align}

Taken together,  
\begin{align}
\frac{\mathcal{Q}(f)}{2k_BT}&=\frac{1}{L^2}\int_0^Ldx\rho_{\text{ss}}^{-1}(x)\times \frac{\langle (I_-(x))^2I_+(x)\rangle_L}{\langle I_-(x)\rangle_L^3}\nonumber\\
&=\varphi_+\varphi_-\left[1+\left(\frac{\langle x\rangle_-}{\varphi_-}-\frac{\langle x\rangle_+}{\varphi_+}+\frac{1}{\varphi_+}\int_0^Ldxe^{\beta V(x)}\left\{\frac{\psi_+(x,0)}{L\varphi_+}-\frac{\Psi_+(x,0)}{L^2\varphi_+\varphi_-}\right\}\right)\beta f+\mathcal{O}(f^2)\right]\nonumber\\
&\times\frac{1}{\varphi_+\varphi_-}\left[1+\left(\frac{2l_o^2-L^2\varphi_+\varphi_-}{2L\varphi_+\varphi_-}\right)\beta f+\mathcal{O}(f^2)\right]\nonumber\\
&=1+\left[\left(\frac{l_o^2-L^2\varphi_+\varphi_-/2+L\langle x\rangle_-\varphi_+-L\langle x\rangle_+\varphi_-}{L\varphi_+\varphi_-}\right)+\frac{1}{\varphi_+}\int_0^Ldxe^{\beta V(x)}\left\{\frac{\psi_+(x,0)}{L\varphi_+}-\frac{\Psi_+(x,0)}{L^2\varphi_+\varphi_-}\right\}\right]\beta f+\cdots\nonumber\\
&=1+\underbrace{\left[\left(\frac{\Psi_+(L,0)-L^2\varphi_+\varphi_-/2}{L\varphi_+\varphi_-}\right)+\frac{1}{\varphi_+}\int_0^Ldxe^{\beta V(x)}\left\{\frac{\psi_+(x,0)}{L\varphi_+}-\frac{\Psi_+(x,0)}{L^2\varphi_+\varphi_-}\right\}\right]}_{=\lambda}\beta f+\cdots.
\end{align}
Now, for $\mathcal{Q}(f)/2k_BT\geq 1$ to be valid, the prefactor of $\beta f$, $\lambda$, should be nonnegative, i.e., $\lambda\geq 0$,  
\begin{align}
\lambda= \frac{1}{\varphi_+}\left(\frac{\int_0^Ldxe^{-\beta V(x)}\int_0^xdye^{\beta V(y)}}{\int_0^Ldxe^{-\beta V(x)}}-\frac{L\varphi_+}{2}+\frac{\int_0^Ldxe^{\beta V(x)}\int^x_0dye^{\beta V(y)}}{\int_0^Ldxe^{\beta V(x)}}-\frac{\int_0^Ldxe^{\beta V(x)}\left(\int_0^xdye^{-\beta V(y)}\int_0^ydze^{\beta V(z)}\right)}{\int_0^Ldxe^{\beta V(x)}\int_0^Ldxe^{-\beta V(x)}}\right)\nonumber\\
\label{eqn:lambda1}
\end{align}
Because $\int_0^Ldxe^{\beta V(x)}\int^x_0dye^{\beta V(y)}=\frac{1}{2}\left(\int_0^Ldxe^{\beta V(x)}\right)^2=\frac{L^2\varphi_+^2}{2}$, 
the second and third terms in Eq.\ref{eqn:lambda1} vanish. 
The numerator of the last term in Eq.\ref{eqn:lambda1} can be rewritten as 
\begin{align}
\int_0^Ldxe^{\beta V(x)}\int_0^xdye^{-\beta V(y)}\int_0^ydze^{\beta V(z)}&=\int_0^Ldxe^{-\beta V(x)}\int_0^xdye^{\beta V(y)}\int_x^Ldze^{\beta V(z)}\nonumber\\
&\leq \int_0^Ldxe^{-\beta V(x)}\int_0^xdye^{\beta V(y)}\int_0^Ldze^{\beta V(z)}
\end{align}
Therefore, $\lambda\geq 0$ is further ensured from the following. 
\begin{align}
\lambda&= \frac{1}{\varphi_+}\left(\frac{\int_0^Ldxe^{-\beta V(x)}\int_0^xdye^{\beta V(y)}}{\int_0^Ldxe^{-\beta V(x)}}-\frac{\int_0^Ldxe^{\beta V(x)}\left(\int_0^xdye^{-\beta V(y)}\int_0^ydze^{\beta V(z)}\right)}{\int_0^Ldxe^{\beta V(x)}\int_0^Ldxe^{-\beta V(x)}}\right)\nonumber\\
&\geq  \frac{1}{\varphi_+}\left(\frac{\int_0^Ldxe^{-\beta V(x)}\int_0^xdye^{\beta V(y)}}{\int_0^Ldxe^{-\beta V(x)}}-\frac{\int_0^Ldxe^{-\beta V(x)}\int_0^xdye^{\beta V(y)}\left(\int_0^Ldze^{\beta V(z)}\right)}{\left(\int_0^Ldxe^{\beta V(x)}\right)\left(\int_0^Ldxe^{-\beta V(x)}\right)}\right)=0
\label{eqn:lambda2}
\end{align}

\subsection{Strong tilt limit:  $V'(x)\ll f$}
Under this condition, $V(x)$ is minor compared to $fx$ term. Thus, we expand $e^{\pm \beta V(x)}$ into  Taylor series. 
\begin{align}
I_-(x)&=\int_0^Ldye^{-\beta(V(x)-V(x+y))}e^{-\beta fy}\nonumber\\
%=e^{-\beta\Phi(x)}\int^{x+L}_xdye^{\beta\Phi(y)}=\int^{L}_0dye^{\beta (V(y+x)-V(x))}e^{-\beta fy}\nonumber\\
&=\int_0^Ldy\left(1+\beta V'(x)y+\frac{1}{2}\left\{\beta^2(V'(x))^2+\beta^{-1} V''(x)\right\}y^2+\cdots\right)e^{-\beta fy}\nonumber\\
&=\frac{1}{\beta f}\left(1+\frac{V'(x)}{f}+\frac{\left\{(V'(x))^2+\beta^{-1} V''(x)\right\}}{f^2}+\cdots\right)
\end{align}
Then 
\begin{align}
\langle I_-(x)\rangle_L=\frac{1}{\beta f}\left[1+\frac{\langle(V'(x))^2\rangle_L}{f^2}+\cdots\right]
\end{align}
\begin{align}
\langle I_-(x)\rangle_L^3=\frac{1}{\beta^3 f^3}\left[1+\frac{3\langle(V'(x))^2\rangle_L}{f^2}+\cdots\right]
\label{eqn:I_3}
\end{align}
where $\langle V'(x)\rangle_L=0$ and $\langle V''(x)\rangle_L=0$ due to the periodicity of $V(x)$. 

Next, 
\begin{align}
I^2_-(x)I_+(x)%&=I_-^2(x)e^{\beta \Phi(x)}\int_{x-L}^xdye^{-\beta\Phi(y)}\nonumber\\
&=I_-^2(x)\int_0^Ldye^{\beta (V(x)-V(x-y))}e^{-\beta fy}\nonumber\\
&=I_-^2(x)\int_0^Ldy\left(1+\beta V'(x)y+\frac{1}{2}\left\{\beta^2(V'(x))^2+\beta^{-1} V''(x)\right\}y^2\cdots \right)e^{-\beta fy}\nonumber\\
&=I_-^2(x)\frac{1}{\beta f}\left(1+\frac{V'(x)}{f}+\frac{(V'(x))^2-\beta^{-1} V''(x)}{f^2}+\cdots\right)
\end{align}
where $\lim_{\beta fL\gg 1}\frac{1}{(\beta f)^{n+1}}\int_0^{\beta fL}d\alpha \alpha^ne^{-\alpha}= \frac{n!}{(\beta f)^{n+1}}$ was used to evaluate the integrals in the second line. 
Then 
\begin{align}
\langle I_-^2(x)I_+(x)\rangle_L&=\frac{1}{L}\int_0^Ldx\frac{1}{\beta^3 f^3}\left(1+\frac{2V'(x)}{f}+\frac{3(V'(x))^2+2\beta^{-1} V''(x)}{f^2}+\cdots\right)\nonumber\\
&\qquad\qquad\qquad\times\left(1+\frac{V'(x)}{f}+\frac{(V'(x))^2-\beta^{-1}V''(x)}{f^2}+\cdots\right)\nonumber\\
&=\frac{1}{\beta^3f^3}\left(1+\frac{3\overbrace{\langle(V'(x))\rangle_L}^{=0}}{f}+\frac{6\langle(V'(x))^2\rangle_L+\beta^{-1}\overbrace{\langle V''(x)\rangle_L}^{=0}}{f^2}+\cdots\right)\nonumber\\
&=\frac{1}{\beta^3f^3}\left(1+\frac{6\langle(V'(x))^2\rangle_L}{f^2}+\cdots\right)
\label{eqn:I2I1}
\end{align}
\end{widetext}
Therefore, by taking the ratio between Eq.\ref{eqn:I2I1} and Eq.\ref{eqn:I_3}, we get the following  for $f\gg V'(x)$ 
\begin{align}
\frac{\langle (I_-(x))^2I_+(x)\rangle_L}{\langle I_+(x)\rangle_L^3}=1+\frac{3\langle (V'(x))^2\rangle_L}{f^3}+\cdots \geq 1. 
\end{align}

Next, from the continuous version of Chebyshev's sum inequality, i.e., for real-valued, integrable functions $X(x)$ and $Y(x)$ satisfying $X'(x)Y'(x)\leq 0$ for $\forall x\in [0,L]$, it holds that 
\begin{widetext} 
\begin{align}
\left(\frac{1}{L}\int_0^LX(x) dx\right)\left(\frac{1}{L}\int_0^LY(x) dx\right)\geq \frac{1}{L}\int_0^LX(x)Y(x) dx. 
\end{align}
\end{widetext}
With the normalization condition $\int_0^L\rho_{\text{ss}}(x,f)dx=1$, and putting $f(x)=\rho_{\text{ss}}(x,f)$ and $g(x)=\rho_{\text{ss}}^{-1}(x,f)$, it follows that   
\begin{align}
\frac{1}{L^2}\int_0^L\rho^{-1}_{\text{ss}}(x,f)dx\geq 1
\end{align}
for all $f$. 

Thus, the following relation can be acquired for $f\gg V'(x)$
\begin{align}
&\mathcal{Q}(f)/2k_BT\nonumber\\
&=\frac{1}{L^2}\int_0^L\rho^{-1}_{\text{ss}}(x,f)dx\times \frac{\langle (I_-(x))^2I_+(x)\rangle_L}{\langle I_+(x)\rangle_L^3}\nonumber\\
&\geq 1+\frac{3\langle (V'(x))^2\rangle_L}{f^3}+\cdots\geq 1.
\end{align} 

\subsection{Near critical force $f\lesssim f_c$}
Near critical force, $f\approx f_c$, the barriers of potential vanishes, and $\Phi(x,f)$ becomes almost flat over the period, $(0,L)$. The shape of the potential would resemble a descending staircase with the period of $L$.
In this case we can approximate $\Phi(x+nL,f)=c-nfL$ for $nL<x<(n+1)L$. 
Then, 
\begin{align}
\langle &I_+(x)\rangle_L=\frac{1}{L}\int^L_0dxe^{\beta\Phi(x,f)}\int_{x-L}^xdye^{-\beta\Phi(y,f)}\nonumber\\
&\approx \frac{1}{L}\int_0^Ldxe^{\beta c}\left(\int_{x-L}^0dye^{-\beta(c+fL)}+\int_{0}^xdye^{-\beta c}\right)\nonumber\\
&=\frac{L}{2}(1+e^{-\beta fL})
\end{align}
and
\begin{widetext}
\begin{align}
\langle [I_+(x)]^2I_-(x)\rangle_L&=\frac{1}{L}\int^L_0dxe^{\beta\Phi(x,f)}\left(\int_{x-L}^xdye^{-\beta\Phi(y,f)}\right)^2\left(\int_x^{x+L}dye^{\beta\Phi(y,f)}\right)\nonumber\\
&\approx \frac{1}{L}\int^L_0dxe^{\beta c}\left(\int_{x-L}^0dye^{-\beta (c+fL)}+\int_{0}^xdye^{-\beta c}\right)^2\left(\int_x^{L}dye^{\beta c}+\int_L^{L+x}dye^{\beta (c-fL)}\right)\nonumber\\
&=\frac{L^3}{12}\left(1+5e^{-\beta fL}+5e^{-2\beta fL}+e^{-3\beta fL}\right)
\end{align}
which leads to  
\begin{align}
\frac{\langle [I_+(x)]^2I_-(x)\rangle_L}{\langle I_+(x)\rangle_L^3}=\frac{2}{3}\frac{\left(1+5e^{-\beta fL}+5e^{-2\beta fL}+e^{-3\beta fL}\right)}{(1+e^{-\beta fL})^3}
\label{eqn:I}
\end{align}
Next, 
\begin{align}
\frac{1}{L^2}\int^L_0dx \rho_{\text{ss}}^{-1}(x,f)&\approx \frac{1}{L^2}\int_0^Ldx\frac{e^{\beta c}\left(\psi_+(L,f)\psi_-(L,f)-f\Omega\int_0^Ldxe^{-\beta c}\int_0^xdye^{\beta c}\right)}{\psi_+(L,f)-\Omega\int_0^xdye^{\beta c}}\nonumber\\
&=\frac{1}{L^2}\int_0^Ldx\frac{e^{\beta c}\left(L^2-\Omega L^2/2\right)}{e^{\beta c}L-\Omega e^{\beta c}x}=\int_0^Ldx\frac{\left(1-\Omega/2\right)}{L-\Omega x}=\frac{1-\Omega/2}{\Omega}\log{\frac{1}{1-\Omega}}\nonumber\\
&=\frac{\beta fL}{2}\frac{1+e^{-\beta fL}}{1-e^{-\beta fL}}
\label{eqn:rho_inv}
\end{align}
Therefore, from Eqs.\ref{eqn:I} and \ref{eqn:rho_inv}, it follows that 
\begin{align}
\frac{\mathcal{Q}(f_c)}{2k_BT}=\frac{\beta f_cL}{3}\frac{\left(1+5e^{-\beta f_cL}+5e^{-2\beta f_cL}+e^{-3\beta f_cL}\right)}{(1+e^{-\beta f_cL})^2(1-e^{-\beta f_cL})}\geq 1.
\label{eqn:Qf}
\end{align}
Eq.\ref{eqn:Qf} is a monotonically increasing function of $\beta f_cL$, greater than 1, and the equality sign is acquired when $\beta f_c L\rightarrow 0$. 
\end{widetext}

%\bibliography{mybib1}

\end{document}